\documentclass[12pt, letterpaper]{JHEP3}

\usepackage{amssymb, amsmath, amsopn, amsthm}
\usepackage{epsfig}

\title{Nongeometric Flux Compactifications}

\author{Jessie Shelton, Washington Taylor, and Brian Wecht \\
{Center for Theoretical Physics} \\ {Massachusetts Institute of Technology} \\
{Cambridge, MA 02139, U.S.A.} \\ {\tt jshelton\ {\rm at}\ mit.edu, wati\
  {\rm at}\ mit.edu, bwecht\ {\rm at}\ mit.edu}}

\abstract{We investigate a simple class of type II string
compactifications which incorporate nongeometric ``fluxes'' in
addition to ``geometric flux'' and the usual $H$-field and R-R fluxes.
These compactifications are nongeometric analogues of the twisted
torus.  We develop T-duality rules for NS-NS geometric and
nongeometric fluxes, which we use to construct a superpotential for
the dimensionally reduced four-dimensional theory.  The resulting
structure is invariant under T-duality, so that the distribution of
vacua in the IIA and IIB theories is identical when nongeometric
fluxes are included.  This gives a concrete framework in which to
investigate the possibility that generic string compactifications may
be nongeometric in any duality frame.  The framework developed in
this paper also provides some concrete hints for how mirror symmetry
can be generalized to compactifications with arbitrary $H$-flux, whose
mirrors are generically nongeometric.}

\preprint{hep-th/0508133, MIT-CTP-3673}

\newcommand{\bZ}{\mathbb{Z}}
\newcommand{\bR}{\mathbb{R}}

\newcommand{\td}{\tilde}

\newcommand{\vol}{{\rm vol}}

\newcommand{\beq}{\begin{equation}}
\newcommand{\eeq}{\end{equation}}

\newcommand{\er}[1]{(\ref{eq:#1})}

\begin{document}
\tableofcontents
\section{Introduction}
\label{sec:}

Since the early days of string theory, it has been clear that there
are many possible ways in which to compactify the various perturbative
superstring and supergravity theories from ten or eleven dimensions to
four space-time dimensions.  For example, compactifying any
ten-dimensional string theory on a Calabi-Yau complex three-fold leads
to a supersymmetric theory of gravity coupled to light fields in the
remaining four macroscopic space-time dimensions.  Moduli
parameterizing the size and shape of the Calabi-Yau appear as massless
scalar fields in the four-dimensional theory.  Understanding and
classifying the range of possible compactifications is an important
part of the program of relating superstring theory to observed
phenomenology and cosmology.  In recent years, compactifications with
topologically quantized fluxes wrapping compact cycles on the
compactification manifold have become a subject of much interest,
following the work of \cite{flux-c, gvw, drs}.  The topological fluxes
produce a 
potential for the scalar moduli, and can thus ``stabilize'' some or
all of the moduli to take specific values \cite{gkp, kst, fp}.  Once
fluxes are added to the system, however, the geometric structure of
the compactification manifold may also become more general.  Recent
work has addressed the generalization to superstring 
compactifications on non-Calabi-Yau geometries \cite{Hitchin,glmw,fmt,su3}.

The goal of the present work is to take the study of flux
compactifications one step further, by including ``compactifications''
which cannot be described by a geometric ten-dimensional space-time
manifold.  It was argued in \cite{kstt} that nongeometric flux
compactifications arise naturally as configurations which are T-dual
to known geometric supersymmetric flux compactifications.  To be
specific, consider for example a compactification on a six-torus $T^6$
with NS-NS 3-form flux $H_{abc}$ on some three-cycle, where indices
$a, b, \ldots\in\{1, 2, \ldots, 6\}$ take values in the compact
directions.  Under a single T-duality, say in direction $a$, this
$H$-flux is mapped to ``geometric flux'' associated with a twist in
the torus topology.  In the presence of this geometric flux, the
metric on the twisted torus acquires a contribution which can be
written as $(dx^a - f^a_{bc}x^c dx^b)^2$, where $f^a_{bc}$ is
integrally quantized and characterizes the ``geometric flux'' of the
compactification.  This kind of ``twisted torus'' has been studied as
a type of Scherk-Schwarz compactification for many years \cite{ss, early-ss-reduction, km},
and is considered in the context of flux compactification in
\cite{glmw,Schulz,dkpz, x1, hre, vz,cfi}, among others.   
Even in the presence of
geometric flux $f^a_{bc }$, however, as was pointed out in
\cite{kstt}, we can perform another T-duality on direction $b$, since
the metric can be chosen to be independent of the coordinate $x^b$.
Carrying out this T-duality explicitly leads to a dual ``torus'',
which is locally geometric, but which cannot be described globally in
terms of a fixed geometry, due to the appearance of a nongeometric
duality transformation in the boundary conditions which patch together
local descriptions of the compactification space.  Nongeometric spaces
of this type were considered in \cite{dh, Hellerman:2002ax, fww, x2, gh}.  In this paper we
label the nongeometric flux resulting from T-duality $T_b$ by
$Q^{ab}_c$, and we determine how fluxes $Q^{ab}_c$ can be incorporated in
the superpotential for a simple T-duality invariant class of IIA and
IIB compactifications.  Note that although we use T-duality to
determine the role of the fluxes in the superpotential, a generic
configuration with fluxes $H, f, Q$ turned on cannot be T-dualized to
a completely geometric compactification.

After performing the second T-duality to a
configuration with nongeometric flux $Q^{ab}_c$, there is no apparent
residual
isometry around direction $c$.  Na\"ively it does not seem that a
further T-duality can be performed.  Nonetheless, we find a structure
suggesting that there is some meaning which can be given to a T-dual
flux of this type, which we label $R^{abc}$.  While we do not have an
explicit presentation of a nongeometric compactification with such
nongeometric fluxes, it seems that this structure should have some
meaning in any background-independent formulation of the theory.
The situation here is analogous to that for R-R fluxes.  The Buscher
rules \cite{Buscher, bho, mo} for T-duality
act on the $p$-form R-R fields $A^{(p)}$, and therefore cannot be used
to explicitly construct configurations with R-R flux $F^{(0)}$, just
as they cannot be used to construct $R^{abc}$, which acts formally as an NS-NS
0-form flux.   In the case of $F^{(0)}$, it is necessary to use
T-duality rules which act directly on the R-R fluxes
\cite{ght,Hassan,kst} to map $F^{(1)} \rightarrow F^{(0)}$. 
Acting on the integrated R-R fluxes, these T-duality rules
take
\begin{equation}
F_{x\alpha_1 \cdots \alpha_p}
\stackrel{T_x}{\longleftrightarrow}
F_{\alpha_1 \cdots \alpha_p} \,.
\label{eq:ft}
\end{equation}
The T-duality rules we construct in this paper for nongeometric
fluxes, which take
\begin{equation}
H_{abc} \stackrel{T_a}{\longleftrightarrow} f^a_{bc}
\stackrel{T_b}{\longleftrightarrow} Q^{ab}_c
\stackrel{T_c}{\longleftrightarrow} R^{abc},
\label{eq:t-chain}
\end{equation}
can be thought of as an extension of (\ref{eq:ft}) to a general class
of integral NS-NS fluxes.  The structure we find here for the simple
toroidal example suggests that in addition to $H$-form flux, both
geometric and nongeometric fluxes can be thought of as additional
algebraic structure added to a given string background.  In the
geometric context, this kind of structure for geometric fluxes as data
which decorate a Calabi-Yau seems to arise naturally when the mirror
of a Calabi-Yau with $H$-flux has a geometric description \cite{glmw,
fmt}.  Another perspective on the geometric and nongeometric structures
arising from the T-dualities in (\ref{eq:t-chain}) is given in
\cite{Mathai}.

The approach taken in this paper is as follows: In Section
\ref{sec:superpotential}, we incorporate nongeometric fluxes by
using
T-duality and coordinate transformations to construct the complete set
of terms which may appear in the superpotential for a class of
compactifications based on a symmetric $T^6 = (T^2)^3$ with all $T^2$
components identical.  The resulting simple polynomial superpotential
subsumes the previously known superpotentials for geometric
compactifications of the IIA and IIB theories, and extends them to a
T-duality invariant form which includes both nongeometric $Q^{ab}_c$
and $R^{abc}$ types of ``fluxes.''  In Section
\ref{sec:interpretation} we give a more detailed discussion of the
interpretation of $Q^{ab}_c$ and $R^{abc}$ fluxes, and provide an
explicit description of ``T-folds'' \cite{x2} with $Q$-structure.  In Section
\ref{sec:constraints} we discuss constraints on the fluxes, which
arise both from tadpole cancellation requirements in the presence of
orientifolds, and from Bianchi-type identities.
Finally, in Section \ref{sec:conclusions} we conclude and discuss some
directions for future research.

\section{Nongeometric fluxes and the superpotential}
\label{sec:superpotential}

In this section we use various dualities to explicitly construct a
simple low-energy effective theory governing a class of geometric and
nongeometric compactifications.  We begin by considering the
compactification of type IIA and type IIB string theory on a torus
$T^6= (T^2)^3$, where each $T^2$ factor represents an identical torus.
We can think of this as a compactification on a $T^6$ with additional
discrete symmetries imposed.  We impose a symmetry under $\bZ_2$ which
reflects the first two 2-tori under (-1, -1, -1, -1, 1, 1) and a
further symmetry under a $\bZ_3$ which rotates the tori $T^2_{(1)}
\rightarrow T^2_{(2)} \rightarrow T^2_{(3)} \rightarrow T^2_{(1)}$.
In the IIB theory we then have a single complex structure modulus
$\tau$ parameterizing the complex structure of the $T^2$, a single
K\"ahler modulus $U$ containing the $C_4 $ modulus and the scale of
the $T^2$ and an axiodilaton $S$.  In the IIA theory which arises
after 3 T-dualities (one on each $T^2$), $\tau$ becomes the K\"ahler
modulus and $U$ becomes the complex structure modulus \cite{vz,cfi}.
Flux compactifications of this type were considered in type IIB in
\cite{kst,fp,x3,dgkt1} and in IIA in \cite{vz,cfi}.  A slightly more
general model was also considered in these papers, where the three
complex structure and K\"ahler moduli are allowed to vary
independently by imposing a second $\bZ_2$ symmetry (1, 1, -1, -1, -1,
-1) instead of the $\bZ_3$ we use here.  This model can be seen as a
special case of that $T^6/\bZ_2^2$ model.  It is straightforward to
generalize the considerations here to that more general model, with
slightly more algebra.

We wish to include various kinds of fluxes on the $T^6$.  When an
orientifold is included to cancel tadpoles,
these
fluxes lead generally to a low-energy effective ${\cal N} = 1$ supergravity
theory in four dimensions which has a superpotential $W(\tau, U, S)$, a K\"ahler
potential $K (\tau, U, S)$, and a resulting potential for the moduli
given by
\begin{equation}
V = e^K \left(\sum_{i,j, = \{\tau, U, S \}} K^{ij} D_i W \overline{D_j
  W} - 3 |W|^2 \right)\,,
\end{equation}
where $K^{ij}$ is the inverse of $K_{ij} =  \partial_i
\bar{\partial_{j}} K$.  The construction of  flux compactifications
with
orientifolds was developed in \cite{gvw,drs}, and applied to the IIB
theory in \cite{gkp} and many subsequent papers, and to the IIA theory
in
\cite{dkpz, vz,cfi,x4, x5, dgkt2}.

An important caveat which must be taken into account when describing
flux compactifications through the dimensionally reduced
four-dimensional theory is that the low-energy four-dimensional
supergravity action is only valid when the moduli acquire masses which
are small compared to those of fields such as higher string modes,
winding modes, and
Kaluza-Klein modes which are neglected in the dimensional reduction.
This issue must be addressed in any study of flux compactifications.
In the class of vacua we consider here, which are not geometric, and
for which ten-dimensional supergravity is not a valid approximation,
this question becomes even more subtle.  For the present, we will
simply describe the superpotential for the four-dimensional
supergravity theory as a function of the degrees of freedom associated
with the original moduli on the torus.  This allows us at least to
characterize the topological features of the nongeometric fluxes in
which we are interested.  We leave a more detailed study of the regime
of validity of the low-energy theory in the presence of nongeometric
fluxes to further work.

The  particular symmetric torus $T^6 = (T^2)^3$ model
we are interested in here was studied in \cite{kst, x3}  and explicitly
solved
in \cite{dgkt1} for IIB
compactifications with R-R and NS-NS form field flux.  In this case the
superpotential is given by
\begin{equation}
W_{{\rm IIB}} = P_1 (\tau) + SP_2 (\tau)
\label{eq:w-IIB}
\end{equation}
where $P_{1, 2} (\tau)$ are cubic polynomials in $\tau$.  The K\"ahler
potential is
\begin{equation}
K = -3\ln (-i (\tau -\bar{\tau}))
 -\ln (-i (S -\bar{S}))
 -3\ln (-i (U -\bar{U})) \,.
\label{eq:Kaehler}
\end{equation}

In \cite{vz,cfi} this model was studied for the IIA theory, where in
addition to NS-NS and R-R form field fluxes, geometric flux was
also allowed.  In this case the K\"ahler potential is again
(\ref{eq:Kaehler}), while the superpotential is
\begin{equation}
W_{{\rm IIA}} = P_1 (\tau) + SP_2 (\tau) + UP_3 (\tau),
\label{eq:w-IIA}
\end{equation}
with $P_1$ again cubic, but with $P_{2, 3}$ now linear in $\tau$.

In order to consider a complete  T-duality invariant family of flux
compactifications we must extend somewhat the nature of allowed
fluxes.  As discussed in the Introduction, we must include not
only geometric fluxes but also some structures we interpret as
``nongeometric fluxes''.
Simply using T-duality and coordinate symmetries, we can proceed to
construct the full duality-invariant superpotential $W$, identifying the
fluxes corresponding to each term in $W$.  We now proceed to directly
present this superpotential, which is one of the main results of this paper,
after which we give a more detailed discussion of how the various
terms are derived through dualities.  The later sections of the paper
discuss the interpretation of the fluxes which appear in this
potential, and constraints on these quantized fluxes.

We claim that the full potential for the symmetric torus in both the
IIA and IIB theories is given by
\begin{equation}
W_{{\rm complete}} = P_1 (\tau) + SP_2 (\tau) + UP_3 (\tau),
\label{eq:w-complete}
\end{equation}
where now all three of $P_{1, 2, 3} (\tau)$ are cubic polynomials.
The coefficients in these polynomials are given in the IIB theory by
(integrally quantized)
NS-NS and R-R fluxes $\bar{H}_{abc}, \bar{F}_{abc}$ 
(denoting the integral number of units of flux of, {\it e.g.},
$F_{abc}$ by $\bar{F}_{abc}$)
and 
also by ``nongeometric''
fluxes $Q^{ab}_c$ (which each can individually arise as the T-dual on
direction $b$ of the geometric
flux  $f^a_{bc}$).  In the IIA theory, the coefficients include 
(integrally quantized) R-R
$p$-form fluxes $F^{(0)}, \bar{F}_{ab}, \bar{F}_{abcd}, \bar{F}_{abcdef}$, as well as
NS-NS 3-form flux $\bar{H}_{abc}$, geometric fluxes $f^a_{bc}$,
nongeometric fluxes $Q^{ab}_c$, and further nongeometric fluxes
$R^{abc}$ (which can individually be seen formally as the T-dual on $c$ of
$Q^{ab}_c$).  In the next section we discuss the interpretation of
these nongeometric fluxes in more detail.  For now, however, we will
simply show how dualities determine which fluxes arise as which
coefficients in the superpotential (\ref{eq:w-complete}).

To make the discussion more explicit, we label coordinates $1, 3, 5$
on the $T^6$ with indices $\alpha, \beta, \gamma$ and coordinates $2,
4, 6$ with indices $i, j, k$.  The IIB torus is taken to have an
O3-plane filling four-dimensional space-time, so that all internal coordinates are
odd under the orientifold reflection $\Omega$.  To get to the IIA
theory we T-dualize on the dimensions $1, 3, 5$, in that order,
so that the resulting
IIA O6-plane extends along  these dimensions with indices $\alpha,
\beta, \gamma$.  In the following table, 
we list the fluxes associated with each term in the superpotential
(\ref{eq:w-complete}) in both IIA and IIB.  
\begin{center}
\begin{tabular}{ || c  || c || c || c ||}
\hline
\hline
Term & IIA flux & IIB flux & integer flux\\
\hline
\hline
$1 $& $ \bar{F}_{\alpha i \beta j \gamma k}$& $ \bar{F}_{ ijk} $& $  a_0 $\\
\hline
$\tau $& $ \bar{F}_{\alpha i \beta j} $& $\bar{F}_{ ij \gamma} $& $   a_1 $\\
\hline
$\tau^2 $& $ \bar{F}_{\alpha i} $& $\bar{F}_{i \beta \gamma} $& $  a_2 $\\
\hline
$\tau^3 $& $ F^{(0)}  $& $\bar{F}_{\alpha \beta \gamma} $& $  a_3 $\\
\hline
$S $& $ \bar{H}_{ijk} $& $ \bar{H}_{ ijk} $& $  b_0$\\
\hline
$U $& $ \bar{H}_{\alpha \beta k} $& $  Q^{\alpha \beta}_k $& $  c_0 $\\
\hline
$S\tau $& $ f^\alpha_{j k} $& $ \bar{H}_{\alpha jk} $& $  b_1 $\\
\hline
$U\tau $& $ f ^ j_{k\alpha}, f^i_{\beta k}, f^\alpha_{\beta \gamma} $& $ 
Q ^{\alpha j}_k, Q^{i \beta}_k, Q^{\beta \gamma}_\alpha $&
$\check{c}_1,  \hat{c}_1,\tilde {c}_1 $\\
\hline
\hline
$S \tau^2 $& $ Q^{\alpha \beta}_k $& $ \bar{H}_{ i \beta \gamma} $& $  b_2 $\\
\hline
$U \tau^2 $& $  Q ^{\gamma i}_\beta, Q^{i \beta}_\gamma, Q^{ij}_k $& $ Q ^{i\beta}_\gamma, Q^{\gamma i}_\beta,
Q^{ij}_k $& $\check{c}_2,  \hat{c}_2,\tilde{c}_2 $\\
\hline
$S\tau^3 $& $  R^{\alpha \beta \gamma} $& $ \bar{H}_{\alpha \beta \gamma} $& $  b_3 $\\
\hline
$U \tau^3 $& $ R^{ij \gamma} $& $  Q^{ij}_{\gamma} $& $c_3 $\\
\hline
\hline
\end{tabular}
\vskip0.5cm
{\small {\bf Table 1}:\,\,Fluxes appearing as coefficients of terms in the superpotential.} 
\end{center}

To be explicit about the index structure in this table, notice
that the orbifold projection we have chosen implies that all objects with
three indices must have one index on each $T^2$; our convention
in the table
is that these indices are ordered cyclically by $T^2$ in the fashion
indicated by the Greek and Latin indices; thus for example $Q^{\alpha
  \beta}_k = Q^{13}_6 = Q^{35}_2 = Q^{51}_4$. The IIA R-R forms
that survive the orbifold projection must have pairs of indices
extending on both dimensions of each
$T^2$ on which there are any indices.  We denote here by $\alpha i,
\beta j, \gamma k$ pairs of indices on the same torus; thus, for
example $\bar{F}_{\alpha i} = \bar{F}_{12}=\bar{F}_{34}=\bar{F}_{56}$.
Additionally, all fluxes in the table are antisymmetric in their upper
indices as well as in their lower indices.  
Note that while $f$ and $R$ fluxes do not appear on the IIB side of
the table, this is a consequence of the fact that all dimensions on
the $T^6$ are odd under the orientifold reflection, and $f$ and $R$
require an even number of odd indices.  In more general orientifolds,
IIB compactifications would also admit $f$ and $R$ fluxes.
  
The resulting full superpotential in the symmetric torus model
is
\begin{eqnarray}
W & = & a_0 - 3a_1 \tau + 3a_2 \tau^2 - a_3 \tau^3\\ & &
 \hspace{0.2in} + S (-b_0 + 3b_1 \tau - 3b_2 \tau^2 + b_3 \tau^3)
\nonumber\\ & &
 \hspace{0.2in} + 3 U (c_0 + (\hat{c}_1 +\check{ c}_1 - \tilde{c}_1)
 \tau - (\hat{c}_2 +\check{ c}_2 - \tilde{c}_2) \tau^2 -c_3
 \tau^3). \nonumber
\end{eqnarray}

At this point let us comment briefly on the nature of the integrally
quantized fluxes appearing in the table.  On the standard geometric
torus, by $\bar{H}_{ijk}, \bar{F}_{\alpha i}, \ldots$ we simply mean
the number of units of $H$ or $F$ integrated over the appropriate
cycle on the torus.  In the presence of geometric flux such as
$f^a_{bc}$ this is slightly more subtle, but can still be made
explicit.  There is a natural basis of Einbeins $\eta^a$ 
satisfying $d\eta^a = f^a_{bc}\eta^b \wedge \eta^c$
\cite{kstt}, which we may use to obtain integrally quantized fluxes such as
$F^{(2)}= \bar{F}_{ab} \eta^a \wedge \eta^b$ for any pair $a, b$, even
when the corresponding R-R flux is not in the cohomology of the
manifold.  For example, turning on units of flux $ \bar{H}_{123}= 1,
\bar{F}_4 = M$ on a standard 4-torus gives a configuration which is
taken by T-duality to $f^1_{23} = 1, \bar{F}_{14} = M$ on the dual
torus with geometric flux.  While the resulting R-R flux $F^{(2)}$ is
not in the cohomology (since the 1-cycle is trivial in homotopy),
there is still a nontrivial integral quantization, as we see from this
explicit T-duality.  There are constraints due to tadpole cancellation
and integrated Bianchi identities which we discuss in Section 4; these
place linear constraints on the R-R fluxes in a fixed geometric
background with nonzero $f^a_{bc}$.  This is presumably how the
K-theory description of R-R charges \cite{K-theory} continues to be
valid in the case with geometrical fluxes (possibly related
considerations appeared in \cite{3}).  We do not have a complete
understanding of how this works in detail, however, particularly when
there is torsion in the cohomology.  In the cases with nongeometric
fluxes $Q^{ab}_c$ and $R^{abc}$, we do not have a specific and
concrete interpretation of the meaning of the quantized fluxes, but
from the approach we take here it seems natural to associate integral
fluxes $\bar{H}_{abc}, \bar{F}_{a_1 \cdots a_p}$ with every cycle on
the original torus; these fluxes appear in the superpotential and are
constrained by the identities we compute in Section 4.  These fluxes
are perhaps best interpreted as some ``dressing'' added to the basic
topological structure of the geometrical torus, in a way which might
naturally generalize to other Calabi-Yau manifolds.

Let us now discuss the detailed derivation of the arrangement of
fluxes in the table above.  As discussed above, the IIB NS-NS and R-R
fluxes appearing in $P_1$ and $P_2$ are already known
\cite{gvw,drs,kst}, as are the IIA NS-NS, R-R, and geometric fluxes
appearing in $P_1$ and the linear parts of $P_2$ and $P_3$
\cite{dkpz, vz, cfi}.  We need to complete the story using T-duality
and coordinate symmetry.  Our conventions for the action of T-duality
on topological R-R and NS-NS fluxes, including the nongeometric
fluxes, is that T-duality removes or adds an index to the first
position of an R-R flux, $T_x:\bar{F}_{i_1 \cdots i_n} \leftrightarrow
\bar{F}_{x i_1 \cdots i_n}$, and acts on NS-NS fluxes by either
raising the first lower index or lowering the last upper index, so for
example $T_b:f^{a}_{bc} \leftrightarrow Q^{ab}_c$.  Since we are only
interested in the action of T-duality on the topological part of the
fluxes, additional moduli-dependent terms which appear in the local
T-duality rules \cite{Buscher,bho,ght,Hassan,kst} are not relevant to this
discussion.

We begin by noting that the first eight lines in the table all contain
known fluxes in the IIA picture.  Each of these fluxes can be
T-dualized directly to IIB.  The R-R fluxes transform to the known IIB
R-R flux coefficients.  The $S$ and $S \tau$ terms are associated with
NS-NS $H$-fluxes and geometric fluxes which transform to the known
$H$-fluxes in the IIB theory.  The $U$ and $U \tau$ terms can
also be T-dualized, and lead to new coefficients in the IIB theory
associated with nongeometric fluxes $Q^{ab}_c$ for various values of
$a, b, c$.  Thus, we can use T-duality to complete the picture for the
first 8 lines in the table.

To proceed further we note that in the IIB model, there is no
geometric distinction between $\alpha$ and $i$ indices, as the
O3-plane does not extend in any of the directions on the $T^6$.  Thus,
by switching the roles of the $\alpha$ and $i$ indices in defining the
complex structure, we exchange $\alpha \leftrightarrow i$, etc..  This
exchange takes $1 \leftrightarrow \tau^3, \tau \leftrightarrow \tau^2$
in the superpotential, and allows us to identify the remaining
$Q^{ab}_c$ coefficients associated with $U \tau^2, U \tau^3$
in the IIB superpotential.

This derivation relies only on duality transformations which can be
performed explicitly (at least for each individual flux), and
therefore is a rigorous demonstration that the IIB theory has the full
superpotential (\ref{eq:w-complete}) with all coefficients in the
cubic polynomials $P_{1, 2, 3} (\tau)$ associated with well-defined
geometric and nongeometric fluxes.  A more detailed discussion of the
nongeometric fluxes  $Q^{ab}_c$ appears in the following section.

To complete the story on the IIA side, we would like to carry our
results from IIB back to IIA using the 3-fold T-duality on the
complete IIB superpotential.  The $S \tau^2$ and $U \tau^2$ terms are
associated in the IIB theory with fluxes which transform to
nongeometric fluxes $Q^{ab}_c$ on the IIA side.  This extends the
superpotential constructed in \cite{vz, cfi} to include nongeometric
fluxes of the $Q$-form.  Note, however, that the terms $S \tau^3$ and
$U \tau^3$ are associated with fluxes in the IIB theory which in the
IIA theory must take the form of a ``T-dual'' on direction $c$ of a
nongeometric flux $Q^{ab}_c$.  We do not know how to carry out such a
transformation explicitly.  It seems, however, that for duality
invariance to be complete, we must introduce a new type of
nongeometric flux in the IIA theory, labeled $R^{abc}$.  We discuss
the possible interpretation of these new topological nongeometric
fluxes in the next section.  Including these fluxes leads to a
superpotential (\ref{eq:w-complete}) which is manifestly T-duality
invariant.  Note that with this complete set of fluxes, not only can
we go from the IIB theory with an O3-plane to the IIA theory with an
O6-plane, but we can also perform a complete 6-fold duality on the IIA
theory.  This duality flips the complex structure on each $T^2$, again
taking $1 \leftrightarrow \tau^3, \tau \leftrightarrow \tau^2$.
Again, for this to be an invariance of $W$ we must include the fluxes
$R^{\alpha \beta \gamma}, R^{ij \gamma}$, which in this case are the
6-fold T-duals of $\bar{H}_{ijk}, \bar{H}_{\alpha \beta k}$.

This completes our construction of the duality-invariant
superpotential for orientifold compactifications of
the generalization of the twisted torus in type IIA
and IIB string theory.  As we have seen, nongeometric fluxes appear as
coefficients of various terms in this superpotential.  In the
following sections we will discuss the interpretation of these
nongeometric fluxes and topological constraints on possible values of
these fluxes.

Given the superpotential we have computed here, it is straightforward
in principle
to choose integral fluxes $a_0, \ldots$ (subject to constraints which
we will discuss in Section 4) and to solve the 
equations of motion.
Given the superpotential (\ref{eq:w-complete}) and the K\"ahler
potential (\ref{eq:Kaehler}), the equations for a supersymmetric
vacuum in the four-dimensional theory are
\begin{equation}
D_\tau W = D_SW = D_UW = 0,
\end{equation}
where
\begin{equation}
D_AW = \partial_AW +  (\partial_AK)W \,.
\end{equation}
For generic flux coefficients in the superpotential (\ref{eq:w-complete}),
the equations for $S$ and $U$ are equivalent to
\begin{eqnarray}
P_1 (\tau) + \bar{S}P_2 (\tau) +UP_3 (\tau)& = &  0 \label{eq:es}\\
P_1 (\tau) + S P_2 (\tau) + \left({2\over 3} U +{1\over 3}\bar{U}\right) P_3 (\tau) & = &  0  \label{eq:eu}\,.
\end{eqnarray}
The remaining ($\tau$) equation is
\begin{equation}
(\tau -\bar{\tau})  \partial_\tau W -3 W = 0  \label{eq:et}\,.
\end{equation}
We defer a detailed analysis of solutions of these equations to a
forthcoming paper \cite{stw-2}, but we will make a few brief comments
here regarding the space of solutions.  Clearly, the space of SUSY
solutions to these equations will include all type IIB and IIA flux
vacua on the geometric symmetric torus, as well as possibly a large
number of vacua with geometric and nongeometric fluxes, which may
generically have no geometric duals.  In \cite{dgkt1}, a family of
supersymmetric IIB vacua on the symmetric torus with $W = 0$ was
identified, corresponding to flux compactifications with vanishing
cosmological constant.  These vacua all have nonvanishing $b_2$ or
$b_3$ and therefore are nongeometric in the IIA picture.  Indeed, it
is easy to see from (\ref{eq:es}-\ref{eq:et}) that there are no
geometric $W = 0$ solutions in the IIA theory, since this would
require either Im $S = 0$ (which is unphysical) or $P_2 =0$.  The
vanishing of $P_2$ in turn implies that either $b_0 = b_1 = 0$ (which
violates the tadpole condition (\ref{eq:tadpole1}), which we derive in
general in Section 4), or $\tau = b_0/3 b_1$ is real (which is again
unphysical).  Thus, admitting nongeometric fluxes expands the set of
Minkowski IIA flux vacua on the symmetric torus from the empty set to
a nonzero set of vacua.  In \cite{cfi} it was argued that when
geometrical fluxes are allowed, there are an infinite family of AdS
vacua in the IIA torus model.  Assuming this result is correct, this
shows that including nongeometric fluxes on the IIB side extends the
finite set of geometric SUSY Minkowski vacua by an infinite set of AdS
vacua.  Beyond these already known results, it seems that generic
nongeometric flux configurations may have no geometric duals, but may
nevertheless lead to acceptable SUSY flux compactifications.  A more
detailed analysis of this issue will appear in \cite{stw-2}.

\section{Interpretation of nongeometric fluxes}
\label{sec:interpretation}

In this section we describe in greater detail the structure of
``T-folds'' with nongeometric $Q$-fluxes, and speculate about the
nature of $R$-fluxes.  So far, our treatment of these fluxes has been
fairly formal.  We have essentially defined a set of T-duality
transformation rules for generalized NS-NS fluxes on the torus through
(\ref{eq:t-chain}), analogous to the T-duality rules for R-R fluxes.
In this section we discuss the nongeometric fluxes $Q^{ab}_c$ and
$R^{abc}$ in greater detail, and discuss when compactifications with such
extra structure can be described at least locally geometrically.  We
also comment on the world-sheet description of compactifications with
nongeometric fluxes.

In order to develop some intuition for the nongeometric fluxes $Q
^{ab}_c$, let us discuss a simple example where only the flux $Q
^{ab}_c$ is present.  We will construct this configuration by step by
step using T-duality, as in \cite{kstt, Schulz}, beginning from a
square three-torus with metric \beq ds ^ 2 = dx ^ 2+dy ^ 2+dz ^ 2 \eeq
and $N$ units of $H$-flux, $\bar{H}_{xyz}=N$. We are free to choose a
gauge where \beq
\label{eq:b-example}
B_{xy} = N z.  \eeq We can think of this configuration as a $T ^ 2$
parameterized by $x$ and $y$, fibered over a circle with coordinate
$z$. The NS-NS degrees of freedom coming from reduction on $T ^ 2_{xy} $ are
the complex structure modulus of the torus, $\tau$, and the K\"ahler
modulus $\rho = B_{xy} + i\vol_{xy}$. The perturbative duality group
of the theory reduced on $T ^ 2_{xy} $ is, up to discrete factors, $SL
(2, \bZ)_\tau \times SL (2, \bZ)_\rho$.  The presence of $N$ units of
$H$-flux is described in this language as a nontrivial monodromy
$\rho\rightarrow\rho + N$ as $z\rightarrow z+1$.

Since it will be useful in our discussion of the nongeometric fluxes,
let us take a moment to develop this $SL(2,\bZ)$ description of
compactification with $H$-flux a little further. This will help in understanding a general 
set of quadratic constraints on the fluxes which we derive in the next section.  In an ordinary
dimensional reduction, we take the fields to be independent of the
coordinates of the compact directions.  In a reduction with $H$-flux,
there is nontrivial coordinate dependence in the gauge potentials.
Dimensional reduction in the presence of flux can thus be understood
as a class of Scherk-Schwarz generalized dimensional reduction
\cite{ss, km}.  From this point of view, the $z$-dependence
of the fields is understood by specifying an $z$-dependent element
$M(z) \in SL(2,\bR)_\rho$ which has the desired monodromy
$\left ( \begin{matrix} 1 & N \cr 0 & 1
\end{matrix}
 \right)$
when $z\rightarrow z + 1$ \cite{llp,hull-98}. 
The theory reduced on the three-torus must be independent of $z$, so in the present example $M(z)$ 
can be at most linear in $z$. Our choice of gauge \er{b-example} for $B$ gives us 
\beq
\label{eq:mz}
M(z)= \left(
\begin{matrix}
1 & Nz \cr
0 & 1
\end{matrix}
\right )_{\rho};
\eeq
other choices of $M(z)$ with the same monodromy are possible and lead to reduced theories
which are equivalent under field redefinitions \cite{hull-98}. The reduction ansatz for an arbitary field $\phi$ is
then $\phi(z)= \left[M(z)\right]_\phi \phi_0$, where
$\left[M(z)\right]_\phi $ is the appropriate representation of
$M(z)$, and $\phi_0$ is a vector in this representation containing the
degrees of freedom analogous to zero-modes in this background.

Let us consider the case where the degrees of freedom are R-R field strengths; this will be useful in understanding how turning on topological NS-NS fluxes affects the topological R-R fluxes.  For illustration, consider the topological fluxes $\bar{F}_{wxy}$ and $\bar{F}_w$ in type IIB.  Here $w$ denotes some compact direction transverse to the three-torus.  These degrees of freedom transform under $SL(2,\bZ)_\rho$ as
\beq
\left(\begin{array}{c} \bar{F}_{w xy}\\\bar{F}_w\end{array}\right)\rightarrow\left(\begin{array}{cc} a & b\\c & d\end{array}\right)\left(\begin{array}{c} \bar{F}_{w xy}\\\bar{F}_w\end{array}\right).
\eeq
Therefore, using the matrix \er{mz} to describe the presence of $H$-flux gives us
\beq
\left(\begin{array}{c} {F}_{w xy}(z)\\ {F}_w(z)\end{array}\right)=\left(\begin{array}{cc} 1 & Nz\\ 0 & 1\end{array}\right)\left(\begin{array}{c} \bar{F}_{w xy}\\\bar{F}_w\end{array}\right),
\eeq
so
\beq
F ^ {(3)} = (\bar{F}_{wxy} +Nz\, \bar{ F}_w) dw\, dx\, dy,
\eeq
from which we can recover the familiar Bianchi identity
\beq 
dF ^ {(3)} =-N\bar{ F}_w \, dw\, dx\, dy\, dz =-\bar{ F} ^{(1)}\wedge\bar{H}.
\eeq
As usual, integrating this equation on the $(w, x, y, z)$-cycle leads to the constraint
\beq
\bar{F}_{[w} \bar{H}_{xyz]} = 0.
\eeq
Note that when other fluxes are present, including geometric and nongeometric fluxes, 
this constraint will acquire other terms.
The point of view we used to obtain this constraint will prove very useful in obtaining the analogous constraint terms in the presence of geometric flux and nongeometric $Q$-flux, as we will demonstrate below.

Following \cite{kstt}, we take this square three-torus with $H$-flux and perform a T-duality on the $x$ direction. This yields a twisted $T^3$ with
$f^x_{yz} = N$, which in this gauge has the metric 
\beq
\label{eq:twisted-torus}
ds^2 =  (dx-Nz dy)^2 + dy^2 + dz^2
\eeq
and $B=0$. The nontrivial monodromy  is now $\tau \rightarrow \tau -N$, realized
by the action of the matrix
 \beq
M(z)= \left(
\begin{matrix}
1 & -Nz \cr
0 & 1
\end{matrix}
\right )_{\tau}
\eeq
on the fields in the theory. 

Consider now the behavior of Ramond-Ramond fluxes in this background. For concreteness,
consider the fluxes $\bar{F}_{wx}$ and $\bar{F}_{wy}$ in type IIA. These transform under $SL(2,\bZ)_{\tau}$ as
\beq
\left(\begin{array}{c}
  \bar{F}_{w y}\\\bar{F}_{wx}\end{array}\right)\rightarrow\left(\begin{array}{cc} a &
    b\\c & d\end{array}\right)\left(\begin{array}{c}
     \bar{ F}_{wy}\\\bar{F}_{wx}\end{array}\right). 
\eeq
Therefore, in this background, we find $F ^ {(2)}$ is appropriately expanded as 
\beq
F ^ {(2)} = (\bar{F}_{wy} - Nz \bar{F}_{wx}) dw\, dy+ \bar{F}_{wx} dw\, dx.
\eeq
This reproduces, from another point of view, the expansion of $F ^{(2)}$ on a twisted torus in 
a basis of globally defined 1-forms $\{\eta^x= d x-Nzdy,  \eta^y=dy\}$, where we define $\bar{F}_{ab} $ through $F ^ {(2)} = \bar{F}_{ab}\eta ^ a\wedge\eta ^ b$
\cite{kstt}. 
The Bianchi identity for $F ^ {(2)}$ is as usual
\beq
d F ^ {(2)} =-N dz\, \bar{F}_{wx} dw\, dy =-\bar{ F}_{wx} f ^ x_{yz} dw\, dy\, dz\,
\eeq
which we may freely integrate over the non-twisted $(w, y, z) $-cycle  to obtain the constraint term 
(in the absence of other fluxes)
\beq
 \bar{F}_{x[w}f^x_{ yz]} = 0.
\eeq

As the metric \er{twisted-torus} does not depend on $y$, we may perform another T-duality in the $y$ direction to  arrive at a $T^3$ with nongeometric flux $Q^{xy}_z = N$. The metric
on this background is 
\beq
ds ^ 2 = {1\over 1+N ^ 2 z ^ 2}\left( dx ^ 2+dy ^ 2\right) + dz^2
\eeq
and the $B$-field becomes
\beq
B_{xy} ={Nz\over 1+N ^ 2 z ^ 2}.
\eeq
In this background, the nontrivial monodromy is
${1\over\rho}\rightarrow{1\over\rho} + N$ as $z\rightarrow z + 1$,
which mixes the metric and the $B$-field of the two-torus.  
Given our particular choice of gauge, the presence of the non-zero $Q$-flux is described by the $SL(2,\bR)$ matrix
\beq
\left(\begin{array}{cc} 1 & 0\\Nz & 1\end{array}\right)_\rho.
\eeq

Now consider the behavior of Ramond-Ramond fields in this background. Returning to type IIB with the field strengths $\bar{F}_{wxy}$ and $\bar{F}_w$ turned on, we can see that in the presence of the $Q $-flux, we must write $F ^ {(1)} $ as
\beq
F ^{(1)} = (\bar{ F}_w +Nz\, \bar{ F}_{wxy}) dw,
\eeq
and therefore we obtain the modified Bianchi identity
\beq 
dF ^ {(1)} = -N \bar{ F}_{w xy} dw\, dz = -Q ^{xy}_z \bar {F}_{xyw} dw\, dz .
\eeq
Further, as there are no nongeometric twists on the
directions $ z$ and $w$, we may regard $ dF ^ {(1)}$ as a
two-form without ambiguity.  Integrating this two-form over the $(z,
w)$ 2-cycle, we find the constraint term (again in the absence of other fluxes)
\beq
\bar{F}_{xy  [w}Q^{xy}_{z]} = 0.
\eeq
This is an example of a new kind of constraint, which will be important to take into account in constructing  nongeometric flux compactifications; we will find the general constraints incorporating this type of term in the next section via T-duality.

There is a natural question that arises at this point: 
Given a nonzero $Q^{xy}_z$,
can one
meaningfully perform another T-duality in the $z$-direction? By
analogy with the above, we would expect that such an
operation raises another
index, producing the object we have denoted $R^{xyz}$. In Section 2, we have
argued for the existence of this object in the IIA theory
by asserting that the
four dimensional superpotential in
IIA must agree through T-duality with
that of the IIB theory, so it would be nice to be
able to see this directly arise from T-dualizing $Q_z^{xy}$. 

The simplest situation in which one would expect the $R^{abc}$ terms
to appear is in a $T^3$ compactification with a single type of flux,
where we would hope to 
have a sequence of T-duality transformations on NS-NS fluxes given by
\beq
\bar{H}_{xyz} \stackrel{T_x}{\longleftrightarrow} f^x_{yz}
\stackrel{T_y}{\longleftrightarrow} Q^{xy}_z
\stackrel{T_z}{\longleftrightarrow} R^{xyz} 
\eeq
As we have discussed above,
the first two steps in this procedure can be implemented while
thinking of the $T^3$ as a $T^2$ bundle over an $S^1$, but this
interpretation breaks down in the final step because of the necessary
$z$-dependence of the metric and $B$-field describing $Q^{xy}_z$. 
We leave the ten-dimensional interpretation of this flux for future
work.  
For practical purposes such as computing flux vacua and the properties
of the low-energy effective action, we can certainly make progress
without an explicit description of these nongeometric $R$-fluxes; all
we need for such computations is the
four-dimensional superpotential which uses these objects. 
Eventually, however, for a full understanding of vacua with 
$R$-fluxes, we need some way of explicitly describing such
compactifications in string theory.

It is possible that the $R^{xyz}$ have no interpretation using
conventional notions of local spacetime, and in this sense are truly
nongeometric.  Studying these objects may help us to better
understand how both geometric and nongeometric structures may emerge
from a fundamental formulation of string theory.  Indeed, we can argue
that a truly background-independent formulation of string theory (such
as string field theory) must include backgrounds with nongeometric
$R$-fluxes as follows: Imagine we have a complete,
background-independent formulation of both type IIA and IIB string
theory.  The formulation of IIA and IIB in a standard toroidal
compactification background with no fluxes must be equivalent.  In the
IIB background there are SUSY flux vacua with the flux $H_{\alpha
\beta \gamma}$ turned on; many explicit examples of such vacua were
described in \cite{kst, dgkt1}.  Such a background should be connected
to the background without fluxes in a nonperturbative fashion in the
complete formulation of the IIB theory.  But the
background-independent descriptions of the IIA and IIB theories must
be equivalent.  Thus, there must be a nonperturbative procedure in the
IIA model for going from the vacuum without fluxes to the vacuum with
$R$-flux.  This indicates that $R$-fluxes must have meaning in a
background-independent formulation of the theory.  Of course, one
might prefer the IIB description without $R$-flux when it is
available, but generically there may be vacua where both the IIA and
IIB descriptions have $R$-flux, in which case a more explicit
description of the nature of these nongeometric fluxes would be highly
desirable.

It is also worth thinking about the worldsheet interpretation of these
spaces. Many nongeometric compactifications that have been studied in
the past \cite{dh, Hellerman:2002ax, fww} have been claimed to be
related to asymmetric orbifolds \cite{Narain:1986qm}. One particular
type of compactification features nontrivial twists by an element of
the U-duality group when going around a circular direction \cite{dh,
fww}. On general grounds, it was shown in \cite{dh} that in such
examples some moduli are fixed by requiring that the vacuum lie at the
minimum of a Scherk-Schwarz potential; this is essentially the same as
saying that the moduli will always lie at fixed points of the
U-duality twist.

It is not, however, necessarily true that the kinds of nongeometric
vacua we present in this paper have asymmetric orbifold
descriptions. We can see from the $T^3$ example above that the
monodromy $1/\rho \rightarrow 1/\rho + N$ has no fixed points, and as
such does not fall into the class of vacua considered in
\cite{dh,fww}. As such, we have no particular reason to believe that
the compactifications we consider here generically have asymmetric
orbifold descriptions.  It may be possible to construct some more
general asymmetric CFT in which U-duality twists are incorporated at
the level of the world-sheet action. Although the $f$-, $Q$-, and $R$-fluxes are all in the NS-NS sector, and thus may have descriptions with nontrivial boundary conditions implemented on a conventional string worldsheet, this would presumably require
developing some new technology to understand fully.

\section{Constraints}
\label{sec:constraints}

In this section we discuss constraints on the fluxes.  These
constraints arise in two closely related ways.  
First, the new fluxes contribute to the tadpole
constraints associated with R-R fields.  Second, 
the new fluxes contribute to the Bianchi identities for the
R-R and NS-NS fields.  In this section we first derive the general R-R
and NS-NS constraints on fluxes using T-duality, and then specialize
to the particular toroidal compactification of interest in this paper.

\subsection{R-R tadpole and Bianchi identity constraints}

In this subsection we discuss the constraints on the fluxes coming
from the Ramond-Ramond tadpole conditions and Bianchi identities.  
The simplest constraint arises  in the IIB theory, where
there is a tadpole for the R-R four-form field $A_4$ arising from the
Chern-Simons term $\int A_4 \wedge H_3 \wedge F_3$.  
$A_4$ is also sourced by local D3-brane and O3-plane contributions.
Integrating this tadpole constraint over a six-dimensional compactification
manifold gives the topological constraint
\beq
 \bar{F}_{[abc} \bar{H}_{def]} + {\rm local} = 0, \label{eq:fh}
\eeq

By repeated application of T-duality, including both geometric fluxes
$f^a_{bc}$ and nongeometric fluxes $Q^{ab}_c$ and $R^{abc}$, in the
absence of local sources we have the R-R constraints

\begin{eqnarray}
\bar{F}_{[abc} \bar{H}_{def]} & = &  0\label{eq:R-R1}\\
\bar{F}_{x [abc}f^x_{ de]}- \bar{F}_{[ab} \bar{H}_{cde]} & = &  0\label{eq:R-R2}\\
\bar{F}_{xy [a b c}Q^{xy}_{d]}
 -3 \bar{F}_{x[ab}f^x_{cd]}- 2\bar{F}_{[a} \bar{H}_{ bcd]} & = &  0\label{eq:R-R3}\\
\bar{F}_{xyz[a b c]} R^{xyz}  -9\bar{F}_{xy [a b}Q^{xy}_{c]}
 -18 \bar{F}_{x[a}f^x_{ bc]}+6F^{(0)}\bar{H}_{[abc]} & = &  0 \label{eq:R-R4}\\
\bar{F}_{xyz[a b]} R^{xyz}  + 6\bar{F}_{xy  [a}Q^{xy}_{b]}
-6\bar{F}_xf^x_{[ab]} & = &
0
\label{eq:R-R5}\\
\bar{F}_{xyza} R^{xyz}  -3\bar{F}_{xy}Q^{xy}_a & = &  0\label{eq:R-R6}\\
\bar{F}_{xyz} R^{xyz} & = &  0 . \label{eq:R-R7}
\end{eqnarray}

If we restrict to the geometric context, these constraints are just
versions of the standard Bianchi identity $(d + H)F = 0$. The individual $FH$, $Ff$, and $FQ$ terms
appearing in these constraints were demonstrated explicitly in the simple example discussed in the previous section.
As mentioned previously, in the presence of fixed geometric fluxes, the constraints 
\er{R-R1}-\er{R-R5}
give linear conditions on integrally quantized fluxes $\bar{F}$ which
may not be in the cohomology.
The extra terms with $Q$'s and $R$'s are additional contributions from
nongeometric fluxes.  While we have derived these constraints from
T-duality on the torus, we expect that there may be a much more
general class of compactifications in which these constraints apply.

In our toroidal compactification, we have a set of O3-planes in the IIB model
which sets the RHS of (\ref{eq:R-R1}) to  16 $ \times\left(\begin{array}{c}
6 \\3
\end{array} \right)^{-1}$, where the last factor comes from the
combinatorial factors associated with $F \wedge H$.  In the IIA model,
the corresponding O6-planes set the RHS of (\ref{eq:R-R4}) to 16
$\times$ 6
when the free indices $a, b, c$ run over the directions $i, j, k$.  In
terms of the integer coefficients $a_0, \ldots,$ the resulting tadpole
constraint is the same in both models and is \beq
\label{eq:tadpole1}
a_0 b_3-3 a_1 b_2+3 a_2b_1-a_3b_0 = 16.
\eeq

There is only one further R-R constraint relevant for our model, which
comes from (\ref{eq:R-R5}) for the IIB model and again from
(\ref{eq:R-R4}) in the IIA model.  This constraint becomes
\beq
\label{eq:tadpole2}
a_0 c_3+a_1 (\check{c}_2 + \hat{c}_2-\tilde{c}_2)
-a_2 (\check{c}_1 + \hat{c}_1 -\tilde{c}_1)-a_3c_0 = 0.
\eeq
All remaining tadpole constraints from (\ref{eq:R-R1}--\ref{eq:R-R7}) are satisfied automatically in the
particular background we are considering here.

\subsection{NS-NS Bianchi identity constraints}

We can carry out a similar analysis of the NS-NS Bianchi identities
from T-duality.
In a geometric compactification, the NS-NS fluxes must satisfy
\beq
f ^ x_{[ab} \bar{H}_{cd]x} = 0,
\eeq
which comes from the Bianchi identity $dH = 0$.  
Using T-duality, we find the set of NS-NS constraints
\begin{eqnarray}
\label{eq:ns-1}
\bar{H}_{x[ab}f^x_{cd]} & = &  0\\
\label{eq:ns-2}
f^a_{x[b}f^x_{cd]} + \bar{H}_{x[bc}Q^{ax}_{d]} & = &  0\\
\label{eq:ns-3}
Q^{[ab]}_xf^x_{[cd]}-4f^{[a}_{x[c}Q^{b]x}_{d]} + \bar{H}_{x [cd]} R^{[ab]x} & = &  0\\
\label{eq:ns-4}
Q^{[ab}_xQ^{c]x}_d + f^{[a}_{xd} R^{bc]x} & = &  0\\
\label{eq:ns-5}
Q^{[ab}_x R^{cd]x} & = &  0.
\end{eqnarray}
Finally, in order for the $f$- and $Q$-fluxes to be individually T-dual to $H$-flux, they must satisfy 
\beq
f ^ x_{xa} = 0 = Q ^{ax}_x.
\eeq

Equations (\ref{eq:ns-1}--\ref{eq:ns-5}) have a
nice interpretation in the four-dimensional effective theory. Ignoring
the R-R fields for the purposes of this discussion, in a reduction on
$T ^ 6$ without flux, the four-dimensional supergravity theory
contains a gauge sector with gauge group $U (1) ^ {12}$ coming from
the 10-dimensional metric and $B$-field.  As noted in \cite{ss} and
developed in \cite{km}, adding geometric NS-NS fluxes
$\bar{H}_{abc}$ and $f ^ a_{bc}$ to the compactification makes the
gauge algebra of the four-dimensional theory nonabelian; the fluxes
appear as structure constants of the gauge algebra.  Denoting
generators descending from 10-dimensional diffeomorphism invariance as
$Z_m$ and generators descending from the 10-dimensional gauge symmetry
of $B$ as $X ^ m$, the Lie algebra of the compactified theory is
\cite{km}
\begin{eqnarray}
{[Z_a, Z_b]} & = & \bar{H}_{abc} X^ c+f ^ c_{ab} Z_c\\
\label{eq:commute-1}
{[Z_a, X ^ b]} & = &-f ^ b_{ac} X ^ c \\ 
\label{eq:commute-2}
{[X ^ a, X ^ b]} & = & 0.
\end{eqnarray}
The Jacobi identities of this algebra then yield the purely geometric portion of the NS-NS constraints (\ref{eq:ns-1}--\ref{eq:ns-2}).  This algebra may be written in a form which is manifestly covariant under the perturbative duality group $O(6,6,\bZ)$ \cite{km, hre}.

By acting on the four-dimensional theory with elements of $O(6, 6,\bZ) $  corresponding to T-duality, we may deduce how to modify this gauge algebra in the presence of nongeometric fluxes $Q, R$.  We find 
 the commutators (\ref{eq:commute-1}--\ref{eq:commute-2}) are modified in the obvious way,
\begin{eqnarray}
{[Z_a, X ^ b] } & = &-f ^ b_{ac} X ^ c +Q ^{bc}_aZ_c\\
{[X ^ a, X ^ b]} & = & Q ^{ab}_c X ^ c +R ^{abc} Z_c.
\end{eqnarray}
The Jacobi identities of this, fully general, algebra now reproduce the full set of NS-NS constraints (\ref{eq:ns-1}--\ref{eq:ns-5}).

When applied to our toroidal compactification, the constraints
(\ref{eq:ns-1}--\ref{eq:ns-5}) lead to a number of conditions on the
integer coefficients $b_0, \ldots$.  The first set of  conditions
arises from \er{ns-2} in IIB and yields
\begin{eqnarray}
\label{eq:bc1}
c_0 b_2-\td{c}_1 b_1+\hat{c}_1 b_1-\check{c}_2 b_0 & = & 0\\
\check{c}_1 b_3-\hat{c}_2 b_2+\td{c}_2 b_2-c_3 b_1 & = & 0
\label{eq:bc2}\\
c_0 b_3-\td{c}_1 b_2+ \hat{c}_1 b_2-\check{c}_2 b_1 & = & 0
\label{eq:bc3}\\
\label{eq:bc4}
\check{c}_1 b_2-\hat{c}_2 b_1+\td{c}_2 b_1-c_3 b_0 & = & 0;
\end{eqnarray}
as well as parallel constraints in which all hats and checks are
switched through $\hat{c}_i \leftrightarrow\check{ c}_i$.  In IIA
these constraints come from, in order, \er{ns-2}, \er{ns-4}, and (the
last two) \er{ns-3}.  
For instance, we obtain \er{bc1} in IIB from setting $a=\beta, b=\gamma, c=j$, and
$d=k$ in \er{ns-2}; the others follow similarly. 
The second set of conditions arises from
\er{ns-4} in IIB and yields
\begin{eqnarray}
c_0\td{c}_2-\check{c}_1 ^ 2+\td{c}_1\hat{c}_1-\hat{c}_2 c_0 & = & 0
\label{eq:cc1}\\
c_3\td{c}_1-\check{c}_2 ^ 2 +\td{c}_2\hat{c}_2-\hat{c}_1 c_3 & = & 0\\
c_3 c_0-\check{c}_2\hat{c}_1 +\td{c}_2\check{c}_1-\hat{c}_1\td{c}_2 & = & 0\\
\hat{c}_2\td{c}_1-\td{c}_1\check{c}_2 +\check{c}_1\hat{c}_2-c_0c_3 & =
& 0
\label{eq:cc4},
\end{eqnarray}
as well as the parallel constraints with hats and checks switched.  In
IIA, these constraints again come from \er{ns-2}, \er{ns-4}, and (the
last two) \er{ns-3}.

We can simplify these conditions significantly by subtracting each
equation from its parallel counterpart with hats and checks switched.
{}From (\ref{eq:cc1}-\ref{eq:cc4}), we find the conditions
\begin{eqnarray}
c_1 \Delta_1 & = &  c_0 \Delta_2\\
c_3 \Delta_1 & = &  c_2 \Delta_2\\
\tilde{c}_2 \Delta_1 & = & - \tilde{c}_1 \Delta_2\\
(2\tilde{c}_2 + \hat{c}_2 + \check{c}_2) \Delta_1 & = &  
(2\tilde{c}_1 + \hat{c}_1 + \check{c}_1) \Delta_2
\end{eqnarray}
where
\begin{eqnarray}
c_1 & = & (\tilde{c}_1 + \hat{c}_1 + \check{c}_1)\\
c_2 & = & (\tilde{c}_2 + \hat{c}_2 + \check{c}_2)\\
\Delta_i & = & \hat{c}_i-\check{c}_i, \;\;\;\;\; i \in\{1, 2\} \,.
\end{eqnarray}
Assuming both $\Delta$'s are nonzero allows us to rewrite equations
(\ref{eq:cc1}-\ref{eq:cc4}) in terms of the three components of
$c_1$.  All 4 equations reduce to the same quadratic
\begin{equation}
3 \tilde{c}_1^2 + 3 \tilde{c}_1 (\hat{c}_1 +\check{ c}_1) +
\hat{c}_1^2 +\check{ c}_1^2 + \hat{c}_1\check{ c}_1 = 0.
\end{equation}
This equation has no real solution for $\tilde{c}_1$ unless $\hat{c}_1
=\check{ c}_1$, so $ \hat{c}_1$ and $\check{c}_1$ can be identified.
A similar argument demonstrates that, even after setting
$\hat{c}_1 =\check{ c}_1$, we must have $\hat{c}_2 =\check{ c}_2$.
Thus, the full set of constraints is just 
(\ref{eq:bc1}-\ref{eq:bc4}) and (\ref{eq:cc1}-\ref{eq:cc4}), with
$\hat{c}_i =\check{ c}_i$.

The equality $\hat{c}_i =\check{ c}_i$ implies a convenient anti-symmetry property of the
$Q^{ab}_c$ and $f ^ a_{ bc}$ in our model.  Given the equality $Q^{i \beta}_k =
Q^{\alpha j}_k$, we may through cyclic permutation of the tori obtain $Q^{\alpha j}_k =
Q^{\beta k}_i = -Q^{k \beta}_i$.  One may show similarly that
$Q^{ab}_c$ is antisymmetric under exchange of any upper and lower index,
provided that both indices are of the same kind (Greek or Latin); the
same is true for $f^a_{bc}$. Note, however, that neither $Q^{ab}_c$
nor $f^a_{bc}$ is fully antisymmetric in all three indices, since we
are not free to exchange Latin and Greek indices.

Given the simplification
$\hat{c}_i =\check{ c}_i$, the constraints
(\ref{eq:bc1}-\ref{eq:bc4}) and (\ref{eq:cc1}-\ref{eq:cc4}) can be
simplified further.  In particular,
(\ref{eq:bc1}) and (\ref{eq:bc4}) become equivalent, and
(\ref{eq:bc2}) and (\ref{eq:bc3}) become equivalent when the
constraints on the $c$'s are imposed.  Details of the
parameterization of solutions to these constraints will be presented
along with solutions of the SUSY vacuum equations in \cite{stw-2}.

We close this section with a brief discussion of S-duality.  The IIB
theory is invariant under an S-duality symmetry which exchanges the
fluxes $F_{abc}$ and $H_{abc}$ (with a change of sign in one
direction), while taking $S \rightarrow -1/S$.  This has the effect in
the superpotential of exchanging the integral flux parameters $a_i
\leftrightarrow b_i$.  We expect that it should be possible to combine
this S-duality transformation with T-duality to get a larger U-duality
group under which our 4D theory is invariant.  The constraint
(\ref{eq:tadpole1}) is indeed invariant under S-duality.  The
remaining constraints, however, provide a puzzle.  The equation
(\ref{eq:tadpole2}) is precisely the sum of the independent $bc$
constraints (\ref{eq:bc3}) and (\ref{eq:bc4}) when $a$ and $b$ are
exchanged.  Thus, the constraints are not obviously incompatible with
S-duality, but also
do not precisely match.  
This apparent mismatch in constraints presumably arises from the fact
that the $Q$'s actually transform nontrivially under S-duality, since
they generically mix the  $B$-field and the metric.
Indeed, the mismatch can be seen directly by noting that in
the $FQ$ term in \er{R-R5}
and the $HQ$ term in \er{ns-2} the free indices
(and the number of constraints) do not match.  
It is clearly crucial to better
understand the
effects of S-duality on nongeometric fluxes. We leave this as an open
question for future work.

\section{Conclusions}
\label{sec:conclusions}

In this paper we have developed a framework for systematically
describing nongeometric NS-NS fluxes in the context of a simple
toroidal compactification of type II string theory.  Like R-R fluxes,
the geometric and nongeometric NS-NS fluxes act in some sense as
$p$-forms on a canonically chosen space-time, here $T^6$, and
transform under T-duality by adding and removing lower indices through
\begin{equation}
H_{abc} \stackrel{T_a}{\longleftrightarrow} f^a_{bc}
\stackrel{T_b}{\longleftrightarrow} Q^{ab}_c
\stackrel{T_c}{\longleftrightarrow} R^{abc}  \,.
\label{eq:t-chain-2}
\end{equation}
While we do not have a complete mathematical description of these
objects, at least on the torus we can take (\ref{eq:t-chain-2}) as a
definition of how these nongeometric fluxes transform.  The $f^a_{bc}$
fluxes correspond to geometrical fluxes defining a ``twisted torus''
\cite{ss,km, dh}.  The $Q^{ab}_c$ fluxes describe compactifications on
locally geometric spaces with nongeometric global boundary conditions,
such as previously discussed in \cite{dh, fww, x2, gh}.  We can
explicitly carry out 
T-duality from $H \rightarrow f \rightarrow Q$ using standard Buscher
T-duality rules, so our discussion here is on well-established
ground.  We cannot, however, use the Buscher rules to T-dualize
$Q^{ab}_c \rightarrow R^{abc} $, just as the Buscher rules on R-R
potentials $A^{( p)}$ cannot lead to a direct construction of the R-R
0-form flux $F^{(0)}$.  In this sense, the last T-duality in
(\ref{eq:t-chain-2}) must at this point be taken as a formal definition.

The need to include nongeometric fluxes of the $R$-type becomes clear
in our construction of the superpotential describing the moduli of the
toroidal compactification to four dimensions.  We have used concrete T-duality
constructions to understand how the $Q$'s extend the superpotential in
the type IIB case, where there are no $R$'s allowed in our particular
orientifold compactification due to parity constraints.  The
consistency of the IIA and IIB pictures then forces us to the
conclusion that the nongeometrical fluxes $R^{abc}$ must be included
on the IIA side.  An important open question is whether these $R$-type
fluxes admit a locally geometric description like the $Q$-type fluxes.

In this paper we have focused on nongeometric fluxes associated with a
toroidal compactification.  It is natural to ask how this structure
generalizes to other Calabi-Yau manifolds.  The structure we have
found indicates that nongeometric fluxes may be thought of as some
additional data which decorates the structure of some particular
Calabi-Yau geometry.  This could naturally lead to a generalization of
mirror symmetry, in which a Calabi-Yau in IIA or IIB decorated with
one set of general NS-NS $H$, geometric, and nongeometric fluxes is
mapped through mirror symmetry to the mirror Calabi-Yau in IIB or IIA
decorated with the dual set of NS-NS fluxes.  In particular, in the
picture of mirror symmetry as T-duality on a toroidal fibration
\cite{syz}, $H$-flux with one leg on the $T^3$ fiber would map to
geometric $f$-flux, $H$-flux with two legs on the $T^3$ fiber would
map to nongeometric $Q$-flux, and $H$-flux wrapping the $T^3$ fiber
would map to $R$-flux on the mirror Calabi-Yau.  The situation where
the mirror of a Calabi-Yau with $H$-flux is geometrical (i.e., the
$H$-flux has 0 or 1 legs on the $T^3$) has recently been described in
detail in \cite{glmw, fmt}, following a suggestion in \cite{Vafa},
using the generalized Calabi-Yau geometry developed by Hitchin; it
would be very interesting to understand whether there is a precise way
of extending that work to the nongeometric context considered here.
The generalized tadpole and integrated Bianchi identities we derived
in this paper should be valid in a more general context than just the
toroidal model considered here, and may provide a good starting point
for the concrete generalization of the picture presented in this
paper.

It would also be nice to understand how S-duality fits into this
framework.  As we have discussed at the end of Section 4, it is
natural to expect that the framework we discuss here should be
invariant under a full U-duality group generated by T-duality and
S-duality transformations.  The constraints we have found on the
geometric and nongeometric fluxes seem compatible with S-duality, but
are not manifestly invariant, so some additional structure may be
needed to form the full U-duality invariant picture.  We leave the
resolution of this question as an outstanding problem for future work.

In this paper we have focused on a set of essentially topological
features of string compactifications characterized by a general set of
NS-NS fluxes.  We have described the interplay between these integral
fluxes and a set of degrees of freedom (the torus moduli) chosen by
considering the light degrees of freedom in the particular background
without fluxes.  As we change fluxes, in different regions of flux
space other stringy degrees of freedom will become light, as
discussed for example in \cite{gp}.  Thus, in many cases the low-energy
effective theory described by the superpotential we have computed
here will not give a complete description of the physics.  This is an
issue with any classification of flux vacua, but is more acute here
where we do not necessarily have tools to assess the validity of the
low-energy theory when all nongeometric fluxes are turned on.  Indeed,
these nongeometric flux compactifications may generically appear at
sub-string scales where the supergravity approximation is not valid;
since these compactifications also have R-R fluxes, and, even in the
locally geometric case, have complicated boundary conditions, we
currently have no way of describing these backgrounds precisely 
using perturbative string theory.  It is
clearly important to understand better how the compactifications we
describe here can be understood in terms of some fundamental
formulation of string/M-theory.  In the
full theory, the fluxes we have described here should be a
useful tool for classifying and understanding string backgrounds.  In
some cases, such as those dual to geometric compactifications in which
the low-energy effective description is valid, we know by duality that the
low-energy effective description given by the superpotential we have
computed in terms of nongeometric fluxes will still be valid.  It is
likely that there are other backgrounds which have no geometric dual,
in which this low-energy description is still valid, though these
backgrounds will be significantly harder to identify.

An obvious application of the formalism developed in this paper is to
classify the full landscape of type II compactifications on tori
with general NS-NS fluxes.  The first step in this program would be to
determine the vacua arising from the superpotential we have computed
here, after which it is necessary to determine corrections to the
classical vacuum, including those from other fields which may become
light as mentioned above.  We have explicitly computed the
superpotential for the simplest model with 3 moduli, as well as all
constraints on the fluxes.  Solutions to this superpotential will
include not only all geometric IIA and IIB flux compactifications in
this class, but also compactifications which involve nongeometric
fluxes either in one or both pictures.  A more detailed analysis of
the solution space for this model is currently underway and will be
reported elsewhere \cite{stw-2}.  Unless there is some unexpected
general obstruction to the solution of the SUSY equations, it may be
possible to use methods developed here to demonstrate conclusively
that generic string vacua are nongeometric, increasing yet further
the size of the enormous haystack known as the ``string landscape'',
in which we hope to find a compactification correctly describing our
world's phenomenology and cosmology.

\begin{center}
\bf{Acknowledgements}
\end{center}
\medskip
We would like to thank Katrin Becker, Oliver DeWolfe, Michael Douglas,
Andrew Frey, Shamit Kachru, David Kutasov, Greg Moore, Li-Sheng Tseng,
and Daniel Waldram for useful conversations, and other participants in
the Summer 2005 Aspen Supercosmology workshop for helpful comments.
Thanks to O.\ De Wolfe, S.\ Kachru, and G.\ Moore for comments on a
preliminary version of this paper.  WT would like to thank National
Taiwan University for hospitality during part of this work, and the
Aspen Center for Physics for hospitality during the completion of this
work, This work was supported by the DOE under contract
\#DE-FC02-94ER40818. BW is additionally supported by NSF grant
PHY-00-96515.

\end{document}